\documentclass[usenatbib]{emulateapj}
\usepackage{graphicx}

\usepackage{apjfonts}

\makeatletter

\newenvironment{inlinefigure}{%
\def\@captype{figure}%
\noindent\begin{minipage}{0.999\linewidth}\begin{center}}
{\end{center}\end{minipage}\smallskip}
\makeatother

\def\keV{ke\kern-0.05emV}
\newcommand{\chandra}{\emph{Chandra}}

\renewcommand{\arcsec}{\ensuremath{''}}

\def\chandra    {{\em Chandra}\/}
\def\rxj1720    {{RXJ1720.1+2638}\/}

\def\2A     {{2A 0335+096}}
\def\ms1455     {{MS1455.0+2232}}
\def\ltsim{\raise 2pt \hbox {$<$} \kern-1.1em \lower 4pt \hbox {$\sim$}}
\begin{document}

\submitted{Accepted for publication ApJL}

\title{Do radio core-halos and cold fronts in non major merging clusters originate from the same gas 
sloshing?}

\author{Pasquale\ Mazzotta\altaffilmark{1,2} and Simona Giacintucci\altaffilmark{2,3}} 
\altaffiltext{1}{Dipartimento di Fisica Universit\'a di Roma ``Tor Vergata'',
Via della ricerca scientifica 1, I-00133 Roma, Italy; mazzotta@roma2.infn.it}
\altaffiltext{2}{Harvard-Smithsonian Center for Astrophysics, 60 Garden St.,
Cambridge, MA 02138}
\altaffiltext{3}{INAF - Istituto di Radioastronomia, via Gobetti 101, I-40129, Bologna, Italy; sgiaci\_s@ira.inaf.it}
\shorttitle{Cold fronts in clusters and radio core-halos}
\shortauthors{MAZZOTTA \& GIACINTUCCI}

\begin{abstract}

We show an interesting correlation between the surface brightness 
and temperature structure of the relaxed clusters \rxj1720 ~ and 
\ms1455 , hosting a pair of cold fronts, and their central 
core--halo radio source. We discuss the possibility that the 
origin of this diffuse radio emission may be strictly connected 
with the gas sloshing mechanism suggested 
to explain the formation of cold fronts in non major
merging clusters. We show that the radiative lifetime 
of the relativistic electrons is much shorter than the timescale 
on which they can be transported from the central galaxy up to the 
radius of the outermost cold front. This strongly indicates that 
the observed diffuse radio emission is likely produced by electrons 
re--accelerated via some kind of turbulence generated within the 
cluster volume limited by the cold fronts during the gas sloshing.

\end{abstract}

\keywords{galaxies: clusters: general --- galaxies: clusters: individual
  (\rxj1720 , \ms1455 ) --- X-rays: galaxies --- cooling flows}

\section{Introduction}\label{par:intro} 

A number of cool core clusters of galaxies host 
very peculiar diffuse radio sources at their 
center characterized by a {\it core--halo} (CH herafter) 
structure (e.g. PKS\,0745--191;  \citealt{1991MNRAS.250..737B}). 
These objects show an amorphous steep spectrum halo 
which surrounds a bright core coincident with 
the dominant galaxy, with no jets on the kpc 
scale. Even though the connection between 
CHs and cool core clusters is well 
established, the origin of these sources 
remains uncertain. \cite{1995ApJ...451..125S} 
argued that they might be produced by 
the interaction between the radio source 
and the thermal gas cooling at the cluster 
center, i.e. in the form of disruption of 
the jets by the high--pressure ambient 
medium and buoyant effects on the radio 
plasma from the disrupted jets. In this Letter 
we show an interesting spatial correlation 
between the CH emission and X-ray 
features in two galaxy clusters 
\rxj1720 ~ ($z=0.159$)  and \ms1455 ~ ($z=0.258$), 
both showing a pair of cold fronts (CF hereafter) in their core. 
We suggest a connection between the origin 
of the CH structure and the
 same 
gas sloshing mechanism responsible for the 
formation of the CFs. 
We use $H_0=70$~km~s$^{-1}$~kpc$^{-1}$, $\Omega=0.27$, and $\Lambda=0.73$.

\section{X-Ray data preparation and analysis}\label{par:data} 

For the purpose of this Letter we provide a short summary 
of the X-ray data preparation and analysis. We refer to two 
companion papers for details (Mazzotta et 
al. in preparation). The analysis of \rxj1720 
~ was performed combining 3 \chandra ~ observations (OBS\_ID=1453, 
3224, and 4631). A single \chandra ~ observation (OBS\_ID=4192) 
was used for MS1455.0+2232. After flare cleaning, the 
useful exposure time is $42508\,{\rm s}$ and $88556\,{\rm s}$ 
for \rxj1720 ~ and \ms1455 , respectively. All images 
are background subtracted and vignetting corrected.
The spectral analysis was carried out in the 0.7--9~keV 
energy band in PI channels. The spectral fitting was 
performed assuming fixed absorption and metallicity 
appropriate for each cluster ($N_H=3.6\times 10^{20}$~cm$^{-2}$; 
$Z=0.39\ Z_\odot$,  and $N_H=3.0\times 10^{20}$~cm$^{-2}$; 
$Z=0.5\ Z_\odot$, for \rxj1720 ~ and \ms1455 , respectively).
The temperature maps were produced using a radially 
increasing Gaussian smoothing of $\sigma=1$~pixel in 
the center up to $\sigma=10$~pixels in the more external 
regions (see \citealt{2001ApJ...551..160V} for details.)

\section{X-ray image and Temperature map}\label{par:image} 

\begin{figure*}
\centerline{\includegraphics[width=1.\linewidth]{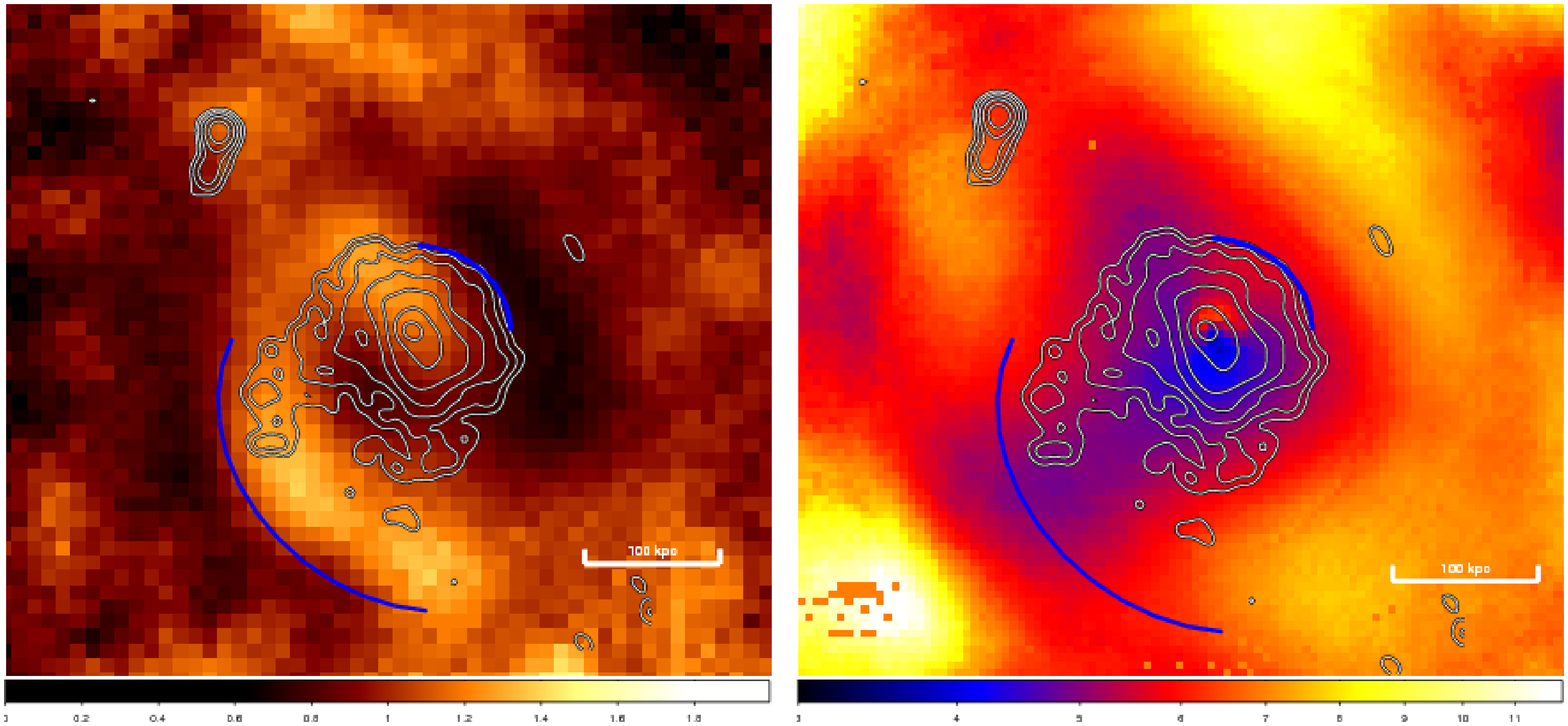}}
\caption{{\it Left panel} -- Ratio \chandra ~ X--ray image of \rxj1720 ~ in 
the [0.5,2.5]~keV band overlaid to the 1.5~GHz VLA--B contours. The X-ray 
ratio image is obtained by dividing the actual cluster image by its radial 
mean values and by smoothing the result using a tophat function with $r=3$~pixels. 
Each pixel corresponds to $4\arcsec$. Blue lines indicate the position of the 
CFs. Radio contours are logarithmically spaced by a factor 2 from 
$0.10$ mJy b$^{-1}$. The radio beam is $5.7\arcsec \times 4.8\arcsec$. 
{\it Right panel} -- X-ray projected temperature map of \rxj1720 ~ in~KeV. 
Contours and lines as in the left panel.}
\label{fig:ratio_rxj1720}
\end{figure*}

\begin{figure*}
\centerline{\includegraphics[width=1.\linewidth]{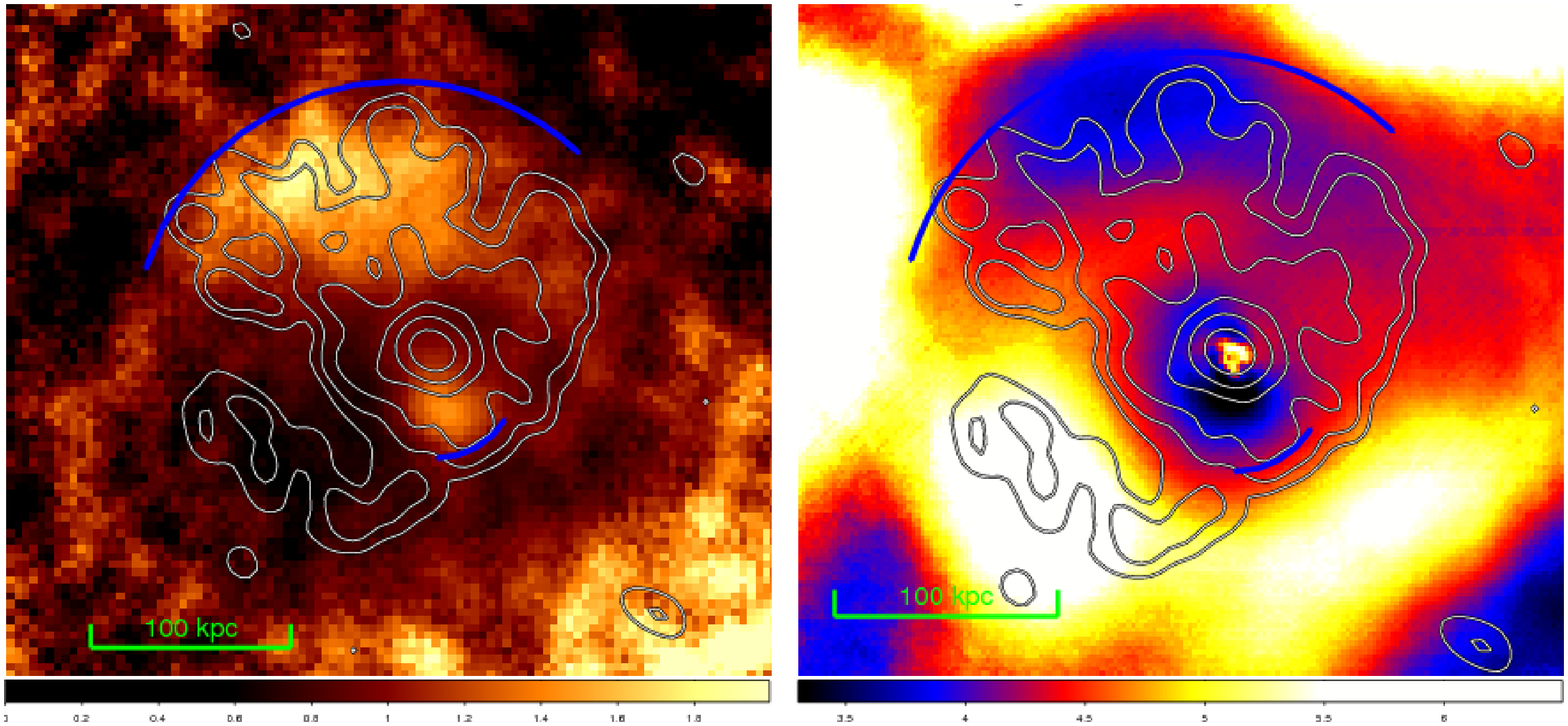}}
\caption{{\it Left panel} -- Ratio \chandra ~ X-ray image of \ms1455 ~ in the 
[0.5,2.5]~keV band overlaid to the GMRT 610~MHz contours. The X-ray ratio image 
is obtained by dividing the actual cluster image by its radial mean values and 
by smoothing the result using a tophat function with $r=3$~pixels. Each pixel 
corresponds to $1\arcsec$. Blue lines indicate the position of the CFs. 
Radio contours are logarithmically spaced by a factor 2
from $0.15$ mJy b$^{-1}$. The radio beam is $5.9\arcsec \times 4.6\arcsec$. 
{\it Right panel} -- X-ray projected temperature map of \ms1455 ~ in~KeV. 
Contours and blue lines as in the left panel.}
\label{fig:fig:ratio_ms1455}
\end{figure*}

Both \rxj1720 ~ and \ms1455 ~ were known to host two 
sharp variations of the surface brightness (edges) 
on opposite side with respect to the cluster center (\citealt{2001ApJ...555..205M}
and \citealt{2001astro.ph..8476M}). It was suggested 
that such features were CFs but no final conclusion was possible
due to the limited statistics available. 
The new deeper observations allow us to disentangle 
the nature of the observed edges. To better highlight 
the spatial structure of the edges, we created a ratio 
image for both clusters by dividing the photon image 
in the $[0.5,2.5]$~keV energy band by its radial 
mean values simply obtained from the radial profile centered in the X-ray peak.
 The resulting image was smoothed using 
a top--hat function with $r=3$~pixels. The ratio image 
and projected temperature map of \rxj1720 ~ are shown 
in the left and right panels of Fig.~\ref{fig:ratio_rxj1720}, 
respectively. Overlaid are the 1.5 GHz radio contours 
(from VLA--B archive data) of the central CH. 
In the left and right panels of Fig.~\ref{fig:fig:ratio_ms1455} 
we show the same images for \ms1455 , with overlaid 
the GMRT 610 MHz contours. 
Figs.~\ref{fig:ratio_rxj1720} and \ref{fig:fig:ratio_ms1455}
clearly show that the edges
appear as part of a more extended spiral--like excess 
structure at the center of both clusters. 
They  seem to confine the CH radio emission. 
Furthermore, the radio structure appears spatially 
correlated with the X--ray excess, following the spiral 
pattern in its more external regions. 

\section{Cold Fronts}\label{par:cold_fronts}

In this Section we demonstrate that all the edges indicated 
by blue lines in Figs.~\ref{fig:ratio_rxj1720} 
and \ref{fig:fig:ratio_ms1455} are CFs. 
We provide a brief description of these features, 
and refer to the forthcoming papers for a detailed 
analysis and discussion. For each edge we extracted 
the surface brightness profile within the relative sector. 
 The centers and position angles for each sector 
are reported in Tab.~\ref{table:1}. In Figs.~\ref{fig:rxj1720_cf} and \ref{fig:ms1455_cf} we show the surface brightness and projected temperature 
profiles of the two sectors of \rxj1720 ~ and \ms1455 , respectively.
 We modeled the edge
using a simple analytic function for the electron density 
($n_e$) and 3D temperature ($T$), both with 
a discontinuity (a jump) at the front:
  
\begin{equation}
  n_{e}=n_0 
\left\{ 
  \begin{array}{rc} 
    D_n(r/r_{jump})^{\alpha_1} & r<r_{jump} \,  \\ 
    (r/r_{jump})^{\alpha_2} & r>r_{jump} \, , \\ 
   \end{array}
   \right . 
\label{eq:1}
\end{equation}
and
\begin{equation}
  T= T_0
\left\{ 
  \begin{array}{lc} 
    1 & r<r_{jump} \,  \\ 
    D_T\left[\frac {1+(r_{jump}/r_T)^2} {1+(r/r_T)^2}\right]^c & r>r_{jump} \, . \\ 
   \end{array}
   \right . 
\label{eq:2}
\end{equation}

Assuming spherical symmetry we projected both 
density and temperature models, and fitted them 
simultaneously to the observed surface brightness 
and temperature profiles. 
For the temperature projection we used the spectroscopic 
like definition $T_{sl}$ introduced by \cite{2004MNRAS.354...10M}. 

\begin{inlinefigure}
\centerline{\includegraphics[width=1.\linewidth]{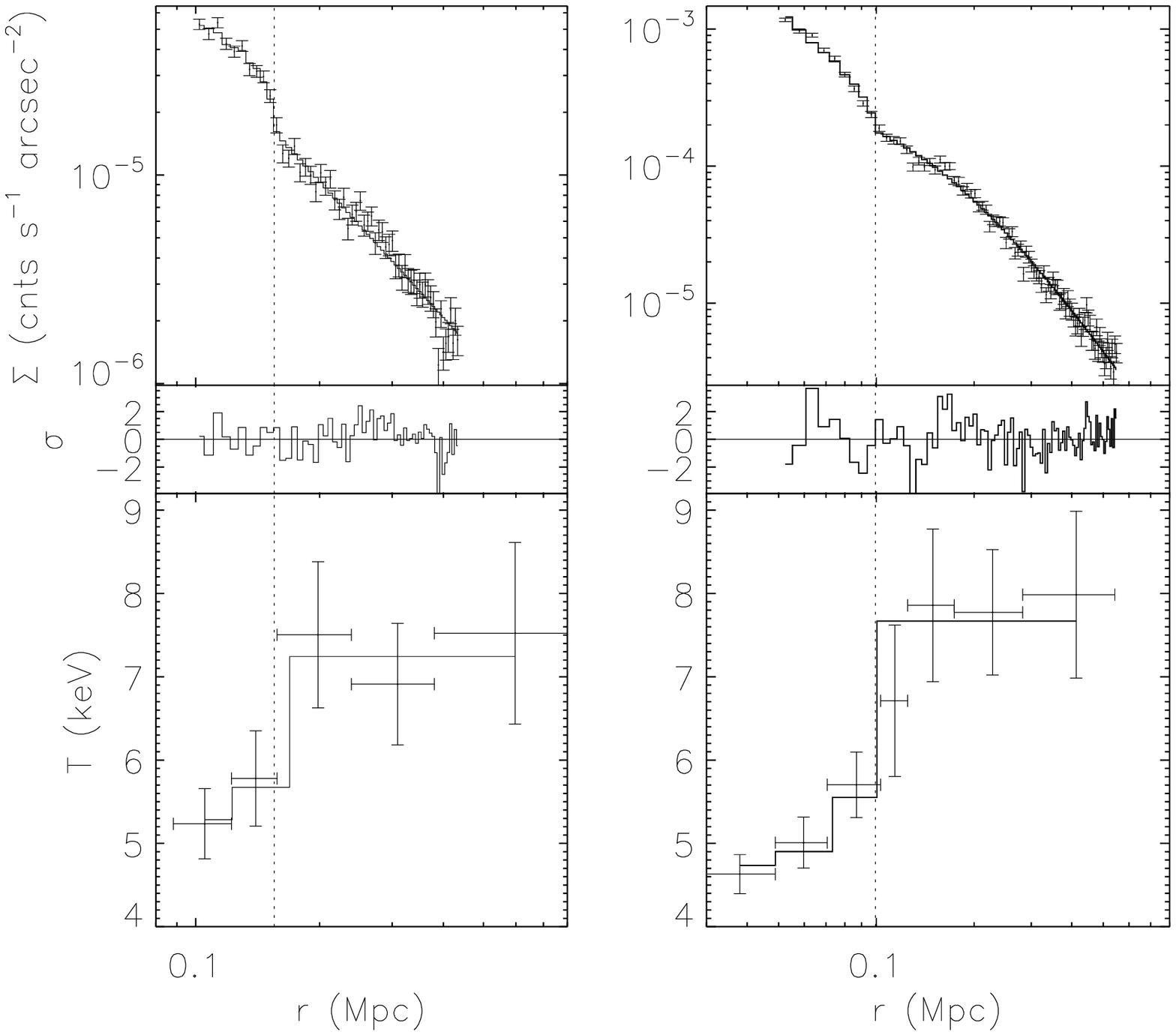}}
\caption{Projected surface brightness and temperature profiles of the cluster 
sectors containing the South-East (left) and North-West (right) 
CFs of \rxj1720 . Vertical dotted lines indicate the positions of the edges. {\it Upper panels}) Surface brightness profiles. 
The continuous histograms are the projected best fit models shown in the upper 
panels of Fig.~\ref{fig:rxj1720_cf_model}. {\it Middle panels}) Residuals of 
the surface brightness profile with respect to the best fit model in sigmas. 
{\it Lower panels}) Projected Temperature profiles. The continues histograms 
are the projected best fit models shown in the middle panels of Fig~\ref{fig:rxj1720_cf_model}}
\label{fig:rxj1720_cf}
\end{inlinefigure}

\begin{inlinefigure}
\centerline{\includegraphics[width=1.\linewidth]{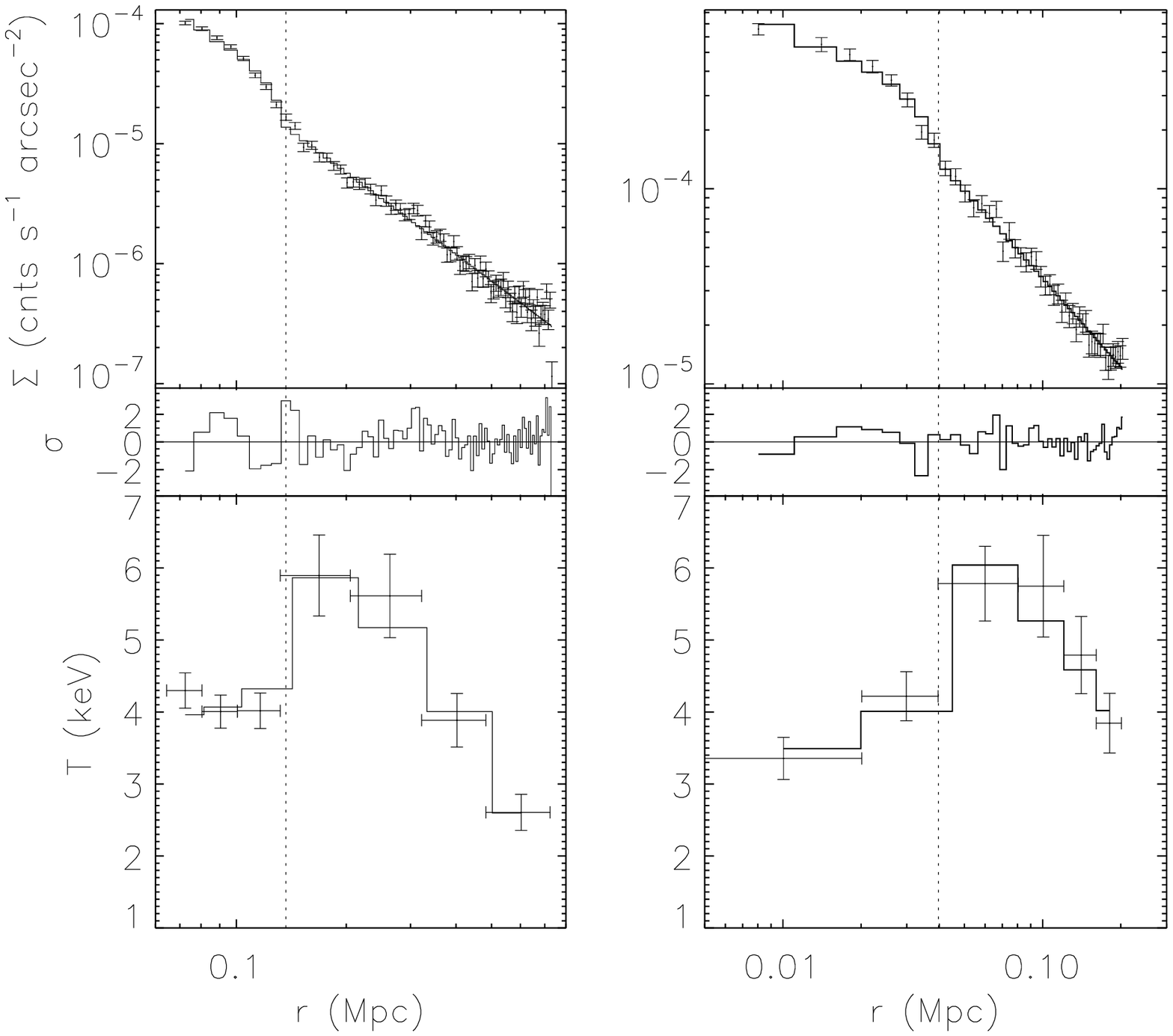}}
\caption{Same as in Fig.~\ref{fig:rxj1720_cf} but for the North-East (left) and 
South-West (right) CFs of \ms1455 . }
\label{fig:ms1455_cf}
\end{inlinefigure}

If Figs.~\ref{fig:rxj1720_cf_model} and \ref{fig:ms1455_cf_model} 
we show the best fit within the 68\% confidence level error 
curves relative to the density, temperature and pressure 
profiles of each edge. We also report as solid line
the best fit models of the surface brightness and temperature
profiles to highlight the good agreement between the fit 
and the data. Furthermore, the middle panels of the same 
figures show the departures in sigmas of the surface brightness 
data with respect to the best fit. 
All the observed edges are CFs. We notice 
that 
the pressure profile is continuous, within the errors, across the front
(Tab.~\ref{table:1}), so the speed 
of the fronts is highly subsonic.  

\begin{inlinefigure}
\centerline{\includegraphics[width=1.\linewidth]{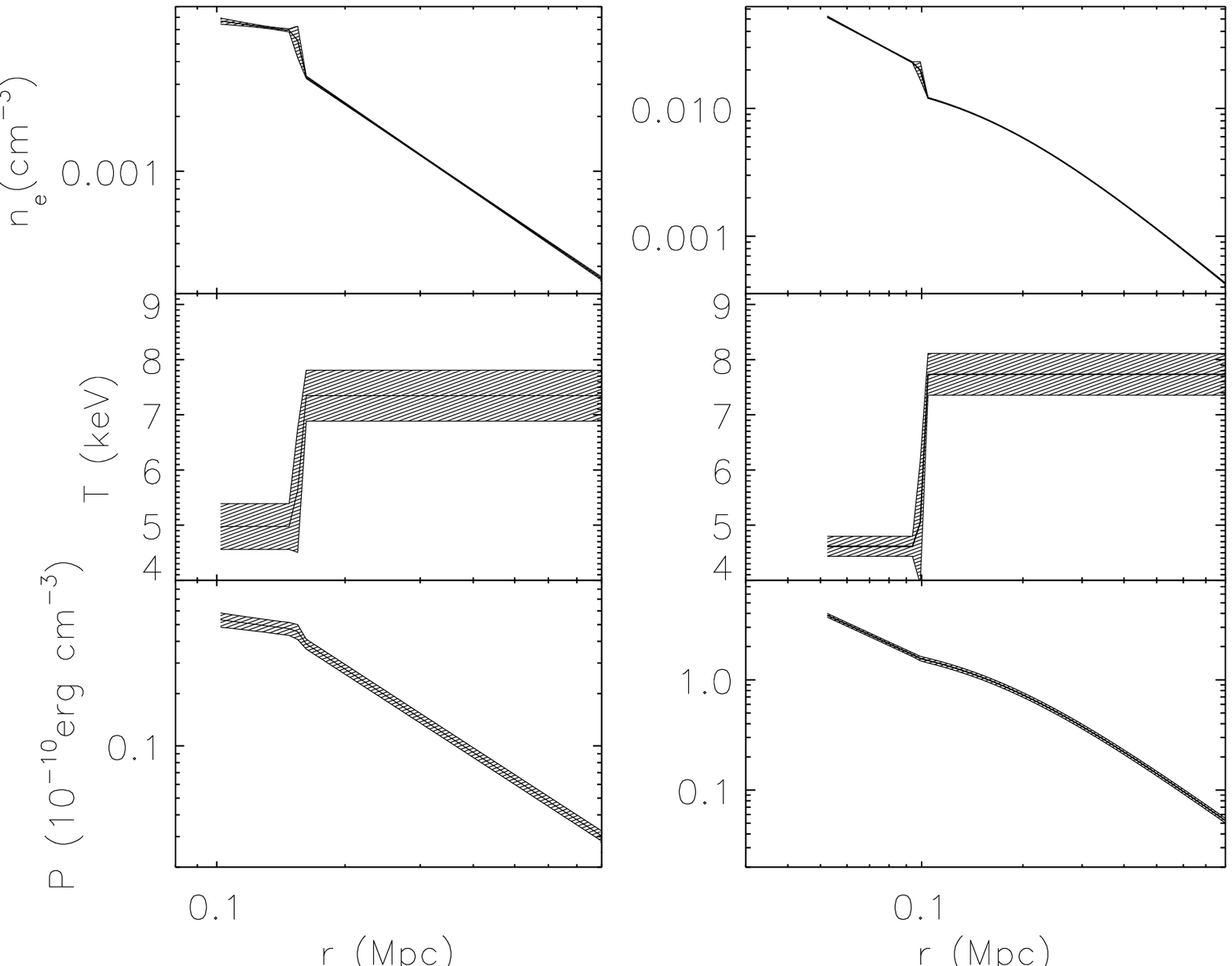}}
\caption{Electron density ($n_e$), temperature ($T$) and thermal pressure ($P$) profiles 
of the sectors containing the South-East (left) and North-West (right) CFs of \rxj1720 . 
The shadow regions indicate the relative 68\% confidence level errors.}
\label{fig:rxj1720_cf_model}
\end{inlinefigure}

\begin{inlinefigure}
\centerline{\includegraphics[width=1.\linewidth]{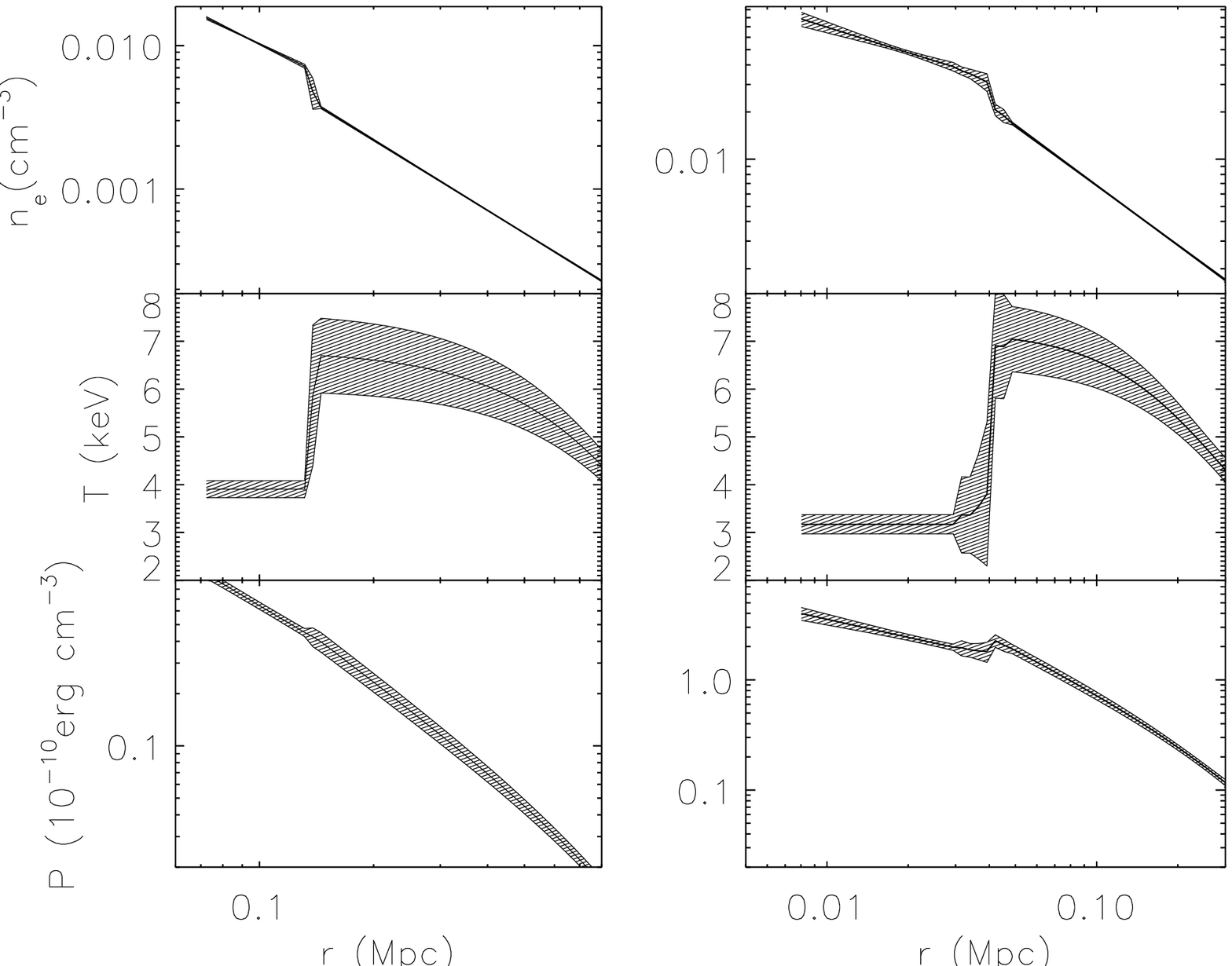}}
\caption{Same as in Fig.~\ref{fig:rxj1720_cf_model} but for the North-East (left) and 
South-West (right) CFs of \ms1455 .}
\label{fig:ms1455_cf_model}
\end{inlinefigure}


\begin{table*}[t]
\centering
 {
\footnotesize
\caption{Parameters of the cold fronts}
\label{table:1}
 \noindent
 \begin{tabular}{ l c c c  c c c c} 
 \hline \hline
 & & & &  & & & \\

 Sector      &$x,y$            & Position angle   & $r_{jump}$ & $D_n$     & $D_T$  & $r_T$   & $c$  \\
             &(J2000)          & (deg){\dag}      &   (Mpc)   &           &        & (Mpc)   &      \\ 
\hline
 & & & &  & & & \\
\rxj1720 ~ N &(17:20:10,+26:37:25) & 2--80              & $0.0994\pm 0.0005$ &  $1.68\pm 0.03$    & $1.68\pm 0.12$ & 0.1 & 0\\
\rxj1720 ~ S &(17:20:10,+26:37:25) & 180--280           & $0.1551\pm 0.0007$ &  $1.66\pm 0.05$    & $1.49\pm 0.16$ & 0.1 & 0 \\
 & & & &  & & & \\
\ms1455 ~ N  &(14:57:15,+22:20:35)&46--163          & $0.1367\pm 0.002$  &  $1.67\pm 0.06$    & $1.72\pm 0.24$ & $1.9 \pm 0.3$ & $2.6 \pm 0.27$\\
\ms1455 ~ S  &(14:57:15,+22:20:30)&270--330         & $0.0397\pm 0.003$  &  $1.50\pm 0.08$    & $2.50\pm 0.30$ & $0.23\pm 0.04$ & $0.5 \pm 0.10$\\
 \hline 
 \multicolumn{8}{l}{\dag Position angles are measured anticlockwise from West}\\ 
 \end{tabular}
 }
 \end{table*}

\section{Discussion and Conclusions}\label{par:discussion}

Using high resolution simulations differet authors
 showed that CFs in the 
core of relaxed galaxy clusters may be the result 
of gas sloshing induced by a minor merger with a 
small dark matter halo occurred within the last few 
Gyrs  (see e.g \citealt{2006ApJ...650..102A} --AM hereafter-- 
and references therein). 
The only condition for the CF to form 
is that the cluster has a steep entropy profile 
(as observed in cool core clusters) and the dark 
matter sub--halo has a non--zero impact parameter. 
When the sub--halo flies through the main 
cluster, it produces a gravitational disturbance that 
pushes the cool gas away from its initial position. 
This makes the gas slosh and two or more CFs 
form on opposite side with respect to the cluster 
center. AM noticed that, when the gas is displaced 
from the center for the first time, it does not fall 
back radially, and the subsequent CF(s) is (are) 
not exactly concentric but combines into a spiral pattern 
(see 1.9--2.1 Gyrs panels in Fig.~7 of AM). As shown 
in \S~\ref{par:image}, both the CF pairs 
in \rxj1720 ~ and \ms1455 , which are clearly two 
distinct non concentric structures, seem to combine 
into a spiral pattern very similarly to what predicted 
by AM. These structures are remarkably similar to 
the simulated ones in the surface brightness map 
(compare left panels of Figs.~\ref{fig:ratio_rxj1720} 
and \ref{fig:fig:ratio_ms1455} with upper--left panel 
of Fig.~19 of AM) as well as in the temperature map 
(compare right panels of Figs.~\ref{fig:ratio_rxj1720} 
and \ref{fig:fig:ratio_ms1455} with lower--left panel 
of Fig.~19 of AM). Thus the X-ray observations of these 
two clusters seems to confirm the prediction of AM for the 
gas sloshing.

What appear to be very interesting is that both clusters host 
a CH radio source. These sources seem to be well confined 
(in projection) within the CF pairs of the respective 
cluster. Furthermore, we noticed a spatial correlation between the
radio emission and the X-ray spiral structure which might suggest 
that the radio emitting electrons are trapped into the same ICM 
flows that produced the CFs. Given the lack of jets on 
the kpc scale, the relativistic electrons need to be transported 
by some other mechanism from the central galaxy up to the radii 
of the outermost CFs. We might speculate that the relativistic 
electrons injected by the central AGN, and trapped into the ICM 
magnetic field, are transported at the CFs radii by the 
motion of the low entropy thermal gas induced by the gas sloshing. 
In fact, the AM simulations show that the CFs are formed 
by the continuous circulation of lower entropy gas that comes from 
the inner core region and falls back soon after (see Fig~7 of AM). 
However, AM show that the speed of this gas circulation is 
relatively low with an upper limit of $v_{max}=500~{\rm km/s}$. 
In both clusters the outermost CF is at a radial 
distance of $\Delta r \approx 150~{\rm kpc}$. Assuming  radial 
transport, we estimate that the minimum time needed for the 
relativistic electrons to be displaced up to the position of the 
outermost front is $t_{min}\ge 3\times 10^{8}\, {\rm yr}$. This time is 
much larger than the radiative lifetime of relativistic electrons
which suffer energy losses mainly due to Inverse Compton with the 
cosmic microwave background photons ($t\approx 10^{7-8}/(1+z)^4\, 
{\rm yr}$; see e.g. \citealt{2001MNRAS.320..365B}). This simple 
calculation suggests that the observed radio emission might be 
produced by re--acceleration of a population of relic electrons,
injected into the ICM by a previous activity of the central AGN.
Due to the subsonic nature of the gas sloshing, we can exclude 
that 
the re--acceleration may be driven by  thermal shock. An 
alternative and plausible mechanisms may be the re--acceleration 
by MHD turbulence, as suggested to explain the cluster 
mini--halos (\citealt{2002A&A...386..456G}) and giant radio halos 
(e.g. \citealt{2004MNRAS.350.1174B}).  These authors claim that, 
to be effective, this re--acceleration mechanism requires 
only a modest level (few per cent) of turbulence.
In fact, 2D simulations
with much higher resolutions have shown that the gas sloshing could
automatically create turbulence in a core (e.g. \citealt{2004ApJ...612L...9F}).

It is worth noting that the CH/CFs correlation 
described in this Letter is not limited to these two clusters 
but seems to be present also in other clusters hosting a CH, 
as e.g. 2A\,0335+096 (\citealt{2003ApJ...596..190M}), 
PKS\,0745--191 (\citealt{1991MNRAS.250..737B}), and A\,2052 
(\citealt{1993ApJ...416...51Z}).
These results calls for further and deeper investigation 
of the radio--X-ray connection in relaxed clusters with 
CFs, which will be addressed in forthcoming papers.

\acknowledgments
 
We thank M. Markevitch, G. Brunetti, T. Venturi, and A. Vikhlinin for 
constructive criticism.  This work was supported by contract ASI-INAF 
I/023/05/0, CXC grants AR6-7015X, GO5-6124X, NASA grant NNG04GK72G 
and by the Smithsonian Institution.


\begin{thebibliography}{10}

\bibitem[{{Ascasibar} \& {Markevitch}(2006)}]{2006ApJ...650..102A}
{Ascasibar}, Y., \& {Markevitch}, M. 2006, \apj, 650, 102

\bibitem[{{Baum} \& {O'Dea}(1991)}]{1991MNRAS.250..737B}
{Baum}, S.~A., \& {O'Dea}, C.~P. 1991, \mnras, 250, 737

\bibitem[{{Brunetti} {et~al.}(2004){Brunetti}, {Blasi}, {Cassano}, \&
  {Gabici}}]{2004MNRAS.350.1174B}
{Brunetti}, G., {Blasi}, P., {Cassano}, R., \& {Gabici}, S. 2004, \mnras, 350,
  1174

\bibitem[{{Brunetti} {et~al.}(2001){Brunetti}, {Setti}, {Feretti}, \&
  {Giovannini}}]{2001MNRAS.320..365B}
{Brunetti}, G., {Setti}, G., {Feretti}, L., \& {Giovannini}, G. 2001, \mnras,
  320, 365

\bibitem[Fujita et al.(2004)]{2004ApJ...612L...9F} Fujita, Y., Matsumoto, 
T., \& Wada, K.\ 2004, \apjl, 612, L9 

\bibitem[{{Gitti} {et~al.}(2002){Gitti}, {Brunetti}, \&
  {Setti}}]{2002A&A...386..456G}
{Gitti}, M., {Brunetti}, G., \& {Setti}, G. 2002, \aap, 386, 456

\bibitem[{{Mazzotta} {et~al.}(2003){Mazzotta}, {Edge}, \&
  {Markevitch}}]{2003ApJ...596..190M}
{Mazzotta}, P., {Edge}, A.~C., \& {Markevitch}, M. 2003, \apj, 596, 190

\bibitem[{{Mazzotta} {et~al.}(2001{\natexlab{a}}){Mazzotta}, {Markevitch},
  {Forman}, {Jones}, {Vikhlinin}, \& {VanSpeybroeck}}]{2001astro.ph..8476M}
{Mazzotta}, P., {Markevitch}, M., {Forman}, W.~R., {Jones}, C., {Vikhlinin},
  A., \& {VanSpeybroeck}, L. 2001{\natexlab{a}}, ArXiv Astrophysics e-prints

\bibitem[{{Mazzotta} {et~al.}(2001{\natexlab{b}}){Mazzotta}, {Markevitch},
  {Vikhlinin}, {Forman}, {David}, \& {VanSpeybroeck}}]{2001ApJ...555..205M}
{Mazzotta}, P., {Markevitch}, M., {Vikhlinin}, A., {Forman}, W.~R., {David},
  L.~P., \& {VanSpeybroeck}, L. 2001{\natexlab{b}}, \apj, 555, 205

\bibitem[{{Mazzotta} {et~al.}(2004){Mazzotta}, {Rasia}, {Moscardini}, \&
  {Tormen}}]{2004MNRAS.354...10M}
{Mazzotta}, P., {Rasia}, E., {Moscardini}, L., \& {Tormen}, G. 2004, \mnras,
  354, 10

\bibitem[{{Sarazin} {et~al.}(1995){Sarazin}, {Baum}, \&
  {O'Dea}}]{1995ApJ...451..125S}
{Sarazin}, C.~L., {Baum}, S.~A., \& {O'Dea}, C.~P. 1995, \apj, 451, 125

\bibitem[{{Vikhlinin} {et~al.}(2001){Vikhlinin}, {Markevitch}, \&
  {Murray}}]{2001ApJ...551..160V}
{Vikhlinin}, A., {Markevitch}, M., \& {Murray}, S.~S. 2001, \apj, 551, 160

\bibitem[{{Zhao} {et~al.}(1993){Zhao}, {Sumi}, {Burns}, \&
  {Duric}}]{1993ApJ...416...51Z}
{Zhao}, J.-H., {Sumi}, D.~M., {Burns}, J.~O., \& {Duric}, N. 1993, \apj, 416,
  51
\end{thebibliography}

\end{document}